\documentclass[lettersize,journal]{IEEEtran}

\usepackage{amsmath,amsfonts}
\usepackage{algorithmic}
\usepackage{algorithm}
\usepackage{array}
\usepackage[caption=false,font=normalsize,labelfont=sf,textfont=sf]{subfig}
\usepackage{textcomp}
\usepackage{stfloats}
\usepackage{url}
\usepackage{verbatim}
\usepackage{graphicx}
\usepackage{mdframed}
\usepackage{cite}
\usepackage{caption}
\usepackage{tcolorbox}
\tcbuselibrary{breakable} 
\usepackage{alltt} 
\usepackage{changepage}
\usepackage{listings}
\usepackage{multirow}
\usepackage{soul}
\usepackage{graphicx}
\usepackage[font=small]{caption} 
\providecommand{\Description}[1]{} 

\begin{document}

\title{Evaluating LLM’s Effectiveness in Detecting and Correcting Test Smells: An Empirical Study}


\author{Enio G. Santana Jr.                     ~\IEEEmembership{UFBA},
        Jander Pereira Santos Junior            ~\IEEEmembership{UFBA},
        Erlon P. Almeida                        ~\IEEEmembership{UFBA},
        Iftekhar Ahmed                          ~\IEEEmembership{UCL},
        Paulo Anselmo da Mota Silveira Neto     ~\IEEEmembership{UFRPE},
        and~Eduardo Santana de Almeida,         ~\IEEEmembership{Senior Member,~IEEE}}
        

\markboth{Journal of \LaTeX\ Class Files,~Vol.~14, No.~8, August~2021}%
{Shell \MakeLowercase{\textit{et al.}}: A Sample Article Using IEEEtran.cls for IEEE Journals}

\IEEEpubid{0000--0000/00\$00.00~\copyright~2021 IEEE}

\maketitle

\begin{abstract}
Test smells indicate poor development practices in test code, reducing maintainability and reliability. While developers often struggle to prevent or refactor these issues, existing tools focus primarily on detection rather than automated refactoring. Large Language Models (LLMs) have shown strong potential in code understanding and transformation, but their ability to both identify and refactor test smells remains underexplored. We evaluated GPT-4-Turbo, LLaMA 3 70B, and Gemini-1.5 Pro on Python and Java test suites, using PyNose and TsDetect for initial smell detection, followed by LLM-driven refactoring. Gemini achieved the highest detection accuracy (74.35\% Python, 80.32\% Java), while LLaMA was lowest. All models could refactor smells, but effectiveness varied, sometimes introducing new smells. Gemini also improved test coverage, unlike GPT-4 and LLaMA, which often reduced it. These results highlight LLMs' potential for automated test smell refactoring, with Gemini as the strongest performer, though challenges remain across languages and smell types.

\end{abstract}


\section{Introduction}
\label{section:introduction}

The development and maintenance of test suites are essential for ensuring software quality and reliability~\cite{spadini_2018, palomba_2016}. Well-structured test suites help verify expected behaviors, detect defects early, and support maintainability, ultimately reducing the cost of software evolution.

However, test smells—sub-optimal coding patterns in test code~\cite{poshyvanyk_2015} - pose significant challenges. These smells make
test code harder to read, understand, and modify~\cite{bavota_2012}, adding unnecessary complexity and obscuring the intent behind test
cases. As a result, they increase maintenance effort and can lead to errors during future development and refactoring.

To address this, automated detection tools have been developed to identify test smells and assist developers in maintaining high-quality test code~\cite{bavota_2012}. By integrating these tools into development workflows (e.g., during code commits), developers can detect and fix smells early, preventing their accumulation over time~\cite{lambiase_2020}. Such tools are especially valuable when developers work under tight deadlines or lack testing expertise.

Meanwhile, advances in Large Language Models (LLMs), such as GPT-4~\cite{gpt4_2024}, LLaMA~\cite{llama_2024}, and Gemini~\cite{gemini_2024}, along with NLP-based tools like CodeBERT and CodeXGLUE~\cite{feng_2020, lu_2021}, have shown great promise in code understanding, generation, and transformation. These models have successfully tackled tasks like code summarization, translation, and repair, suggesting that LLMs could be leveraged to detect and refactor test smells—an area that remains underexplored in research. This led us to answer our first research question, which is:

\noindent\textbf{RQ1:} Are LLMs capable of detecting test smells in test code? To answer this question, we conducted an empirical study evaluating the capabilities of GPT-4, Gemini, and LLaMA in detecting test smells in Python and Java\footnote{These languages were selected due to their popularity and widespread adoption across various domains \cite{github_2022}.}. Our study comprised over 1,400 identification tasks across 15 types of test smells in two programming languages. 

While identifying test smells is an important first step, detection alone is not sufficient to support developers in maintaining high-quality test suites. Developers often lack the time or expertise to address identified smells without additional guidance, and as a result, tools that only detect smells without offering solutions are less likely to be adopted in practice. To be truly useful, tools need to not only flag issues but also assist in automatically refactoring or suggesting concrete fixes. In this context, LLMs hold strong potential, as they have already shown impressive capabilities in automated code refactoring~\cite{saberi_2023, shi_2023} and code generation \cite{An_Empirical_Comparison} 
in non-test code scenarios. Given these successes, we posit LLMs should be able to extend their capabilities to the domain of test code and test smell refactoring, helping developers resolve test smells efficiently and encouraging broader tool adoption in real-world development workflows. This led us to answer our second research question, which is: 

\noindent \textbf{RQ2:} How effective are LLMs in refactoring identified test smells? To answer this question, we analyzed 450 refactoring tasks across 15 types of test smells in two programming languages using GPT-4, Gemini, and LLaMA. Our results indicate that all models generally performed better in Java than in Python. While Gemini exhibited substantial gains in statement coverage for Java projects, GPT-4 delivered the most consistent performance overall, with the least negative impact on missing statement coverage, especially in Python projects.

Overall, this paper makes the following contributions:

\begin{itemize}
    \item We conduct a systematic evaluation of the effectiveness of LLMs, GPT-4, LLaMA, and Gemini, in identify and refactoring test smells in Python and Java test suites.

    \item We integrate traditional detection tools (PyNose and TsDetect) with LLM-based refactoring, enabling a hybrid workflow that 
    combines static analysis and AI-driven code transformation.

    \item We provide a comprehensive analysis of the persistence of test smells after refactoring and assess the impact of automated 
    refactoring on test coverage, offering a multifaceted view of refactoring effectiveness.

    \item  We identify limitations in LLM-based refactoring, including the introduction of new test smells, highlighting current challenges and opportunities for improving the robustness of AI-assisted test code transformation.

    \item To foster replication and future research, we make all experimental artifacts, datasets, and analysis scripts available on our project website \cite{LLMsTestSmellsSupp}. 
\end{itemize}

\section{Related Work}
\label{section:related_work}

This section describes related work on identifying and refactoring test smells, as well as studies that highlight the 
performance of using LLMs models in Software Tetsing.


\subsection{LLMs in Software Testing}
Due to recent developments in LLM, researchers have started investigating the applicability of LLM in Software Engineering. Xinyi et al. \cite{xinyi_2024} conducted a systematic review of 395 papers on LLM applications across the software development lifecycle (SDLC), showing that while 58.37\% of studies focused on development tasks, only 10.30\% addressed quality assurance, with common challenges including overfitting, scalability, and ambiguous code generation. Complementing this, Wang et al. \cite{wang_ieeeTrans_2024} reviewed 102 studies specifically on LLMs in software testing, identifying test case generation and program repair as primary use cases and highlighting the importance of prompt engineering and model fine-tuning. 

Schäfer et al. \cite{Schäfer_ieeeTrans_2024} introduced TestPilot, an LLM-based JavaScript unit test generator that achieved 70.2\% median statement coverage, 19 percentage points higher than existing techniques, while reducing test redundancy. The study concluded that LLMs can effectively generate unit tests without additional training, emphasizing the role of prompt design, model size, and training data in performance. Yaraghi et al. \cite{Yaraghi_ieee} proposed TaRGET, a fine-tuned LLM for automated test repair, reaching 66.1\% exact-match accuracy on 45,373 broken tests. Wang et al. \cite{xinyi_2024} investigated the use of LLMs, including Codex, Codegen, and ChatGPT, to generate edge-case programs for fuzzing deep learning libraries like PyTorch and TensorFlow. Their approach, FuzzGPT, uses historical bug-triggering programs to prime LLMs, outperforming traditional fuzzers by achieving higher code coverage and detecting more bugs. FuzzGPT identified 76 bugs, with 61 confirmed (49 of which were previously unknown), showcasing its ability to uncover edge-case scenarios not present in training data.

Xie et al. \cite{xie2023chatunitest} conducted an evaluation on ChatUnitTest, a tool using GPT-3 to generate unit tests within a validation and repair framework, assessing its performance using correctness and test coverage metrics. Siddiq et al. \cite{siddiq2024empirical} investigated four models, including GPT-3.5-Turbo, for test code generation, evaluating outcomes 
based on correctness, coverage, and test smells. Their findings identified that LLM-generated tests often contain 
test smells, raising concerns about their quality. 

Despite these advances, no prior work has systematically evaluated LLMs' effectiveness in identifying and refactoring test smells. Existing studies on LLMs in software quality assurance have primarily focused on test case generation, automated debugging, and code review \cite{hou2024}, but little attention has been given to their role in improving test maintainability through smell identification and refactoring. Our study aims to fill that gap in research.


\subsection{Non-LLM Test Smell Detection Tools}


Researchers have developed various test smell detection tools, primarily for Python and Java~\cite{ahmed_2021, tempyufba, handling_test_smells_in_python, 10371608, pytest_2022_tt, trend_analysis_test_smell_dblp, antony2024:detecting_smells_python}. Among them, PyNose\cite{ahmed_2021} is a specialized tool for Python, adapting 17 language-agnostic test smells and introducing a Python-specific smell, SuboptimalAssert. It has demonstrated high accuracy, with a precision of 94\% and recall of 95.8\%. For Java, TsDetect\cite{peruma_2020} identifies 19 test smells in JUnit-based test suites using predefined detection rules, producing detailed reports with detected smells, locations, and refactoring suggestions. It achieves precision between 85\% and 100\% and recall between 90\% and 100\%, making it a reliable tool. 

While traditional test smell detection tools like TsDetect~\cite{peruma_2020} and PyNose~\cite{ahmed_2021} achieve high precision and recall, they are often limited by language specificity, heavy reliance on handcrafted rules and limited adaptability to evolving coding practices. To address these limitations, our work empirically evaluates state-of-the-art LLMs: GPT-4, Gemini, and LLaMA, for detecting and refactoring test smells in Python and Java. By comparing their performance against specialized tools, we explore whether LLMs can offer flexible, rule-free alternatives for automated test smell management while maintaining competitive accuracy.
\section{Research Design}
\label{section:research_design}

\begin{figure*}[ht]
    \centering
    \includegraphics[width=1\linewidth]{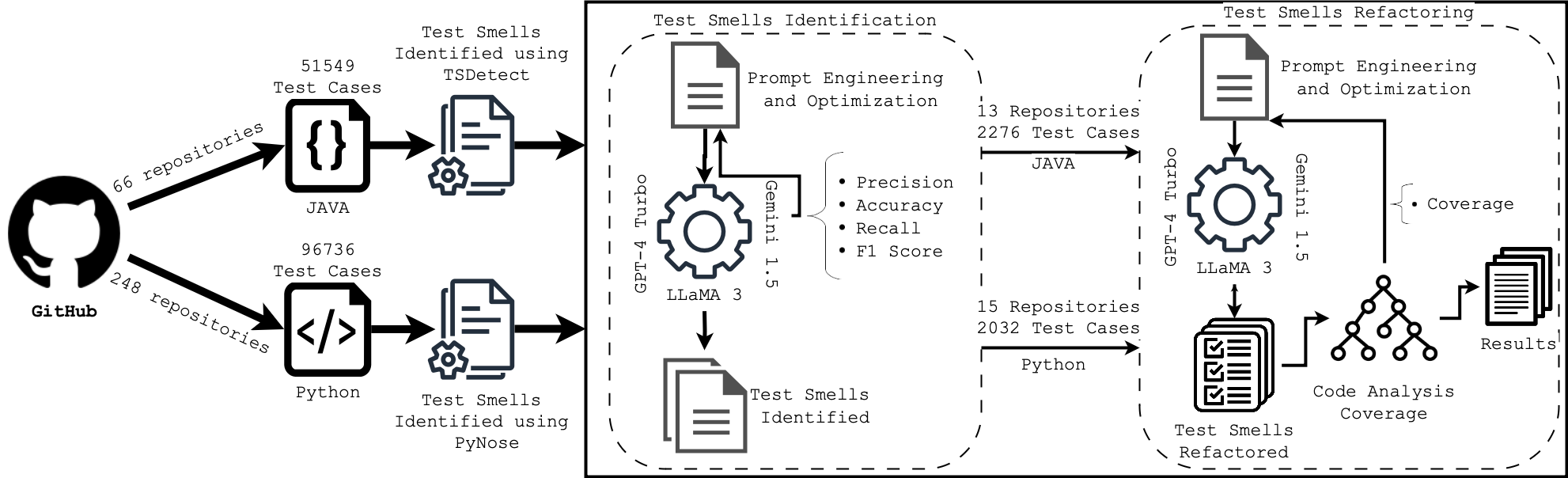}
    \caption{Workflow for Test Smell Identification and Refactoring.}
    \Description{Workflow for Test Smell Identification and Refactoring.}
    \label{fig:research-design1}
\end{figure*}

This section outlines the research design of our study, which follows a structured two-step workflow: (i) test smell identification and (ii) test smell refactoring, as shown in Figure~\ref{fig:research-design1}. Initially, we employed TsDetect (for Java) and PyNose (for Python) as baselines to detect test smells in a curated set of open-source repositories. We then assessed the ability of three state-of-the-art LLMs: GPT-4 (GPT-4-Turbo), Gemini (Gemini-1.5 Pro), and LLaMA (LLaMA 3 70B), to detect and refactor these smells. 
Additionally, we explored prompt engineering strategies to optimize LLM performance across both workflows. Our study is guided by the following research question (RQ):


\medskip
\begin{mdframed}[backgroundcolor=white, linewidth=1pt, linecolor=black]
    RQ1: Are LLMs capable of detecting test smells in test code?
\end{mdframed}
\medskip

Identifying test smells is crucial for maintaining high-quality test code, yet developers often depend on manual analysis \cite{yang_2021, lucia_2016}, which is labor-intensive and prone to mistakes. Automating this process can improve both efficiency and accuracy, reducing the effort needed for test maintenance\cite{santana_2024}. Given LLMs' strong capabilities in code understanding and generation, exploring their potential for test smell detection could enable more effective developer support tools. This motivates our second research question (RQ):


\medskip
\begin{mdframed}[backgroundcolor=white, linewidth=1pt, linecolor=black]
    RQ2: How effective are LLMs in refactoring test code to eliminate identified test smells?
\end{mdframed}
\medskip


While automated test smell refactoring using LLMs holds promise for improving test code quality, it can also introduce unintended side effects. These include the accidental introduction of new test smells, removal of test smells beyond those targeted, and unexpected modifications to test coverage--an essential aspect of test reliability that should remain stable. Such side effects can undermine the goal of refactoring by making tests harder to maintain or less effective. To systematically investigate the impact of LLM-based refactoring on test code quality and reliability, we propose the following sub-research questions (RQs):


\begin{itemize}
    \item \textbf{RQ 2.1:} Which test smell, if any, is unintentionally removed during refactoring?
    \item \textbf{RQ 2.2:} Does the removal of test smells unintentionally introduce new ones?
    \item \textbf{RQ 2.3:} Does the removal of test smells unintentionally impact test coverage?
\end{itemize} 



The following subsections outline the research design, detailing the methodologies and decisions that guided our study. 

\subsection{Dataset Selection}


In this section, we describe the dataset selection processes, including the criteria used by \cite{ahmed_2021} for Python and the methodology from \cite{pontillo_2024} for Java. To identify test smells using machine learning, \cite{pontillo_2024} analyzed 66 open-source Java projects from the International Dataset of Flaky Tests (IDoFT), ensuring diversity and external validity. From 51,549 collected test cases, a stratified sample of 9,633 tests was selected with a 99\% confidence level and a 1\% margin of error, ensuring representativeness \cite{Amoroso:2024}.


For Python projects, \cite{ahmed_2021} utilized three datasets—primary (450 projects), secondary (239 projects), and validation (8 projects)—to ensure robustness and reliability. GitHub repositories collected in July 2020 were processed using PGA-create, applying strict criteria: over 1,000 commits, at least 10 contributors, and a minimum of two years of active development with recent updates. After filtering, 248 primary dataset projects were retained, yielding 9,158 test files and 96,736 test cases for analysis. The dataset for this project is available at \cite{ahmed_2021}.

We selected these datasets for our experiment because the test smell detection tools used to validate the LLM results, specifically 
\textit{PyNose} \cite{ahmed_2021} and \textit{tsDetect} \cite{peruma_2020}, served as oracles. The complete datasets were used for the Test Smell identification task, comprising 66 Java repositories with 51,549 test cases and 248 Python repositories containing 96,736 test cases. For the Test Smell Refactoring task, only projects with all dependencies in which the coverage analysis tools, \textit{PyTest-cov}\footnote{https://pytest-cov.readthedocs.io/en/latest/readme.html} and \textit{JaCoCo}\footnote{https://www.jacoco.org/}, could successfully execute were considered. This filtering step was necessary because the tools rely on dynamic analysis to generate the coverage report. After filtering, this resulted in 13 Java projects (2,276 test cases) and 15 Python projects (2,032 test cases). Coverage reports from refactored tests were then compared against their original versions to assess the impact of refactoring.


\subsection{The Large Language Models}

To assess the effectiveness of LLMs in detecting and refactoring test smells, we conducted the tasks with three state-of-the-art models: 
GPT-4 \cite{gpt4_2024}, LLaMA \cite{llama_2024}, and Gemini \cite{gemini_2024}. GPT-4 and Gemini were accessed via their respective APIs, while LLaMA was deployed locally using Ollama and accessed through an API interface. We selected that models as they represent state-of-the-art LLMs from both proprietary (GPT-4, Gemini) and open-source (LLaMA) ecosystems. These models are being adopted and consistently rank among the top performers in general-purpose reasoning and code-related benchmarks, offering a representative and balanced comparison\footnote{https://www.ucl.ac.uk/news/headlines/2025/jan/ai-chatbots-still-cant-accurately-answer-high-level-history-questions-study}.


We designed our approach to address token constraints by limiting the combined token count of each prompt and test class. This was
done to avoid introducing artificial variability in cases where the correctly refactored class is too large to fit within the response token window.




We evaluated three state-of-the-art LLMs in our experiments. For \textit{GPT}\footnote{\url{https://openai.com/index/gpt-4/}}, we used the standard version of \texttt{GPT-4} accessed via OpenAI's official API, which provides flexible integration, adjustable parameters such as temperature and token limits, and seamless deployment across different environments. For \textit{LLaMA}\footnote{\url{https://ollama.com/library/llama3:70b}}, we employed the standard 70B model running on a local on-premises server via the \texttt{Ollama} platform, which offers a secure and customizable setup to manage and evaluate large language models in controlled environments. For \textit{Gemini}\footnote{\url{https://gemini.google.com/}}, we leveraged the 130B model version through Google DeepMind’s official API, which supports complex reasoning tasks and advanced interactions.

To ensure fairness and reproducibility across models, we kept all parameters at their default settings, except for temperature, which was fixed at 0 for all experiments. This choice aimed to minimize randomness and variability in the responses, enabling consistent and comparable outputs across all LLMs.

\subsection{Test Smell Identification}

Based on the demonstrated effectiveness in the literature, we selected PyNose \cite{ahmed_2021} and TsDetect \cite{peruma_2020}, which have shown high precision and recall in identifying test smells in Python and Java, respectively. The test smells identification process was fully automated using Command Line Interface (CLI) to execute the mentioned tools. Then, we sampled each kind of test smell detected, adopting 95\% confidence level and 5\% margin of error. For each group of sampled data, we also selected an equal number of test code that the detection tool marked as not containing that smell. Next, various prompt techniques (see Table \ref{tab:prompt_techniques}) were applied to determine the most effective method for identifying test smells.

We based our prompt structure on established best practices from prior literature\cite{the_prompt_report}. Our prompt design was organized into three key components: 
\begin{itemize}
    \item {\textbf{Context and Instructions}}: The general definition of what a test smell is, the definition of the specific test smell being targeted, and instructions for the LLM to perform the task (Test smell detection or refactoring).
    \item {\textbf{Complementary Prompting Techniques}}: We experimented adding multiple different prompt techniques to find the best performing prompt template.
    \item {\textbf{User prompt}}: We provide the final command and the specific code to use.
\end{itemize}
Figure \ref{fig:prompt_template} is one example of the final prompt templates we created. The full list of prompts is available on our project website.

\begin{table*}[ht]
  
   \small
   \centering
   \begin{tabular}{|c|p{13cm}|}
    \hline
   \textbf{Prompt Technique}  & \textbf{Description} \\ 
    \hline
   Persona                        & Assigning a specific role or persona to the language model to influence output style and content. \cite{white2023prompt} \\
    \hline
   Zero-shot learning              & Providing the model with direct task instructions without prior examples.                         \cite{zeroshot_techprompt_paper}\\
    \hline
   Few-shot learning                & Offering the model one or more input-output examples to guide text generation.                    \cite{fewshot_learner_techprompt_paper}\\
    \hline
   Chain-of-Thoughts               & Decomposing complex problems into a sequence of intermediate steps for more detailed and interpretable problem-solving. \cite{CoT_prompt_paper}
   \\
   \hline
   Zero-Shot Chain-of-Thoughts (adapted)         & It uses a specific way of providing step-by-step reasoning in a conversational format, different from the CoT technique, which provides a series of numbered steps. \cite{zhao-etal-2024-enhancing-zero}\\
    \hline
   \end{tabular}
    \caption{Prompt Techniques used in our Prompt Templates.} 
   \label{tab:prompt_techniques}
\end{table*}


\refstepcounter{figure}\label{fig:prompt_template}
\begin{tcolorbox}[
    title=Example of Test Smell Detection Prompt for Java,
    colback=white,
    width=\linewidth, 
    colframe=blue!50!black,
    fonttitle=\bfseries,
    boxrule=1pt,
    arc=3mm,
]
\ttfamily\small
\textbf{System prompt}\par
\begin{adjustwidth}{2em}{0em}
You are a software quality assurance engineer specializing in Java. Your task is to assure test code quality by detecting test smells.
\end{adjustwidth}\par

\textbf{Definition of test smells:}\par
\begin{adjustwidth}{2em}{0em}
Test smells represent sub-optimal design or implementation solutions applied when developing test cases.
\end{adjustwidth}\par

\textbf{Definition of Assertion Roulette:}\par
\begin{adjustwidth}{2em}{0em}
Occurs when a test method has multiple non-documented assertions.
\end{adjustwidth}\par

\textbf{Follow the steps below to identify this smell within test code}\par
\begin{adjustwidth}{2em}{0em}
\textbf{i)} Analyze the test code provided;\par
\textbf{ii)} According to the Assertion Roulette definition, check if this smell is present in the provided code;\par
\textbf{iii)} Your output should be ‘YES’ if Assertion Roulette is present, and ‘NO’ if it is not present.
\end{adjustwidth}\par

\textbf{User Prompt}\par
\begin{adjustwidth}{2em}{0em}
analyze test code below:\newline
\textbf{\{Placeholder\}}
\end{adjustwidth}

\end{tcolorbox}
\noindent\textbf{Figure \thefigure. Example of Test Smell Detection Prompt for Java.}\newline

To identify the most effective prompting strategy, we evaluated accuracy, precision, F1-score, and recall (see Table~\ref{tab:comparison-techniques}). In cases where multiple prompts tied on these performance metrics, we selected the prompt with the lowest token count to optimize efficiency. To conduct this evaluation, we sampled code snippets containing both test smells and clean cases (i.e., without smells), and applied each LLM to these samples, asking them to detect both the target test smell and any other potential smells. This process was repeated iteratively until prompt saturation was reached—defined as the point where additional iterations did not yield significant improvement. Saturation was achieved after five iterations per language, resulting in a total of ten distinct prompt versions across Python and Java.

For each iteration, we compared the LLMs' outputs against the results produced by PyNose and TsDetect, which served as oracle tools during this prompt selection phase. Once the best-performing prompt for each language was identified, we applied it to the entire sampled test smell dataset, and again compared the outcomes against the oracle tools to assess final performance. All automation scripts for running these evaluations and detailed information about the sampled datasets are publicly available on our project website.



 \begin{table}[ht]
    \footnotesize
    \begin{tabular}{|p{0.2cm}|c|p{2cm}|c|c|c|c|}
        \hline
        \textbf{} & \textbf{Versions}     & \textbf{Techniques}      & \textbf{ACC} & \textbf{PREC.} & \textbf{REC} & \textbf{F1} \\ \hline
        \multirow{5}{*}{\rotatebox{90}{PYTHON}} 
        
                            & v0 & CoT\cite{CoT_prompt_paper}, 1-Shot\cite{fewshot_learner_techprompt_paper}       & 0.783         & 0.745         & 0.727             & 0.732         \\ \cline{2-7} 
                            & v1 & CoT\cite{CoT_prompt_paper}, 2-Shots\cite{fewshot_learner_techprompt_paper}             & 0.756         & 0.775         & 0.607             & 0.672         \\ \cline{2-7} 
                            & v2 & CoT\cite{CoT_prompt_paper}                 & 0.786         & 0.800         & \textbf{0.647}    & \textbf{0.711}\\ \cline{2-7} 
                            & v3 & ZS-CoT\cite{zhao-etal-2024-enhancing-zero}              & 0.781         & 0.816         & 0.520             & 0.609         \\ \cline{2-7} 
                            & v4 & ZS-CoT\cite{zhao-etal-2024-enhancing-zero}, 1-Shot\cite{fewshot_learner_techprompt_paper}    & \textbf{0.790}& \textbf{0.825}& 0.620             & 0.706         \\ \hline
        \multirow{5}{*}{\rotatebox{90}{JAVA}} 
                            & v0 & CoT\cite{CoT_prompt_paper}, 1-Shot\cite{fewshot_learner_techprompt_paper}       & 0.763         & 0.789         & 0.659         & 0.715         \\ \cline{2-7} 
                            & v1 & CoT\cite{CoT_prompt_paper}, 2-Shots\cite{fewshot_learner_techprompt_paper}             & 0.748         & \textbf{0.806}& 0.607         & 0.682         \\ \cline{2-7} 
                            & v2 & CoT\cite{CoT_prompt_paper}                 & \textbf{0.784}& 0.765         & \textbf{0.766}& \textbf{0.763}\\ \cline{2-7} 
                            & v3 & ZS\cite{zeroshot_techprompt_paper}                  & 0.766         & 0.789         & 0.681         & 0.724         \\ \cline{2-7} 
                            & v4 & ZS-CoT\cite{zhao-etal-2024-enhancing-zero}              & 0.747         & 0.834         & 0.517         & 0.634         \\ \hline
    \end{tabular}
    \caption{Comparative results from prompt versions for Identification task}
    \label{tab:comparison-techniques}
\end{table}

\begin{table}[ht]
    \centering
    \footnotesize
    \begin{tabular}{|c|c|c|c|c|c|}
        \hline
        \textbf{} & \textbf{Versions} & \textbf{Techniques} & \textbf{Direct} & \textbf{Collateral} & \textbf{LOC} \\ \hline
        \multirow{5}{*}{\rotatebox{90}{PYTHON}} 
        
            & v0 & CoT\cite{CoT_prompt_paper}, FS\cite{fewshot_learner_techprompt_paper}       & 477          & 107           & -16585        \\ \cline{2-6} 
            & v1 & CoT\cite{CoT_prompt_paper}, FS\cite{fewshot_learner_techprompt_paper}       & \textbf{477} & 220           & \textbf{-6326}\\ \cline{2-6} 
            & v2 & ZS\cite{zeroshot_techprompt_paper}           & 467          & 32            & -43549        \\ \cline{2-6} 
            & v3 & ZS-CoT\cite{zhao-etal-2024-enhancing-zero} + Hint      & 438          & \textbf{-120} & -17839        \\ \cline{2-6} 
            & v4 & ZS-CoT\cite{zhao-etal-2024-enhancing-zero},       & 468          & -8            & -8685         \\ \hline
        \multirow{5}{*}{\rotatebox{90}{JAVA}} 
        
            & v0 & CoT\cite{CoT_prompt_paper}, FS\cite{fewshot_learner_techprompt_paper}       & 361              & 553           & 10380         \\ \cline{2-6} 
            & v1 & CoT\cite{CoT_prompt_paper}, FS\cite{fewshot_learner_techprompt_paper}       & 314              & 446           & 5064          \\ \cline{2-6} 
            & v2 & ZS\cite{zeroshot_techprompt_paper}           & \textbf{394}     & 264           & \textbf{24141}\\ \cline{2-6} 
            & v3 & ZS-CoT\cite{zhao-etal-2024-enhancing-zero}      & 346              & \textbf{52}   & 19561         \\ \cline{2-6} 
            & v4 & ZS-CoT\cite{zhao-etal-2024-enhancing-zero}, FS\cite{fewshot_learner_techprompt_paper}    & 362              & 640           & 22903         \\ \hline
    \end{tabular}
    \caption{Comparative results from prompt versions in refactoring task.}
    \label{tab:prompt-comparison}
\end{table}

\subsection{Test Smell Refactoring}
The refactoring process was also automated. First, the repositories were filtered based on the availability of dependencies since 
the coverage tools perform dynamic analysis to generate the coverage report. As mentioned before, Table  
\ref{tab:prompt_techniques} describes the prompt techniques used for the refactoring task. The number of test smells refactored was 
used to select the most efficient prompt. In cases where the number of removed and added smells, both those explicitly targeted for
refactoring (direct) and those unintentionally introduced (collateral), resulted in a tie between prompt versions, the prompt with the largest coverage of lines and statements or the smallest number of missing statements was selected. Saturation was achieved after five prompt versions for each language. The results from each iteration were compared with the outcomes from the coverage report, which served as oracles for the prompt technique selection (see Table \ref{tab:prompt-comparison}). Once the best prompts were selected, they were applied to the 13 Java projects, and the 15 Python projects, with samples reaching 320 test cases for the most prevalent test smells. The results were compared (original version vs. refactored version) with the coverage reports.

\section{Results}
\label{section:results}


\subsection{RQ1: Are LLMs capable of detecting the presence/absence of test smells in test code?}

Table~\ref{tab:det_smell_pct} summarizes the identification rates of the three LLMs, categorized by programming language (Python/Java). Each LLM was prompted with the same code snippet for each test smell to ensure a fair comparison.

The results demonstrate that Gemini consistently performed well across both Python and Java, often outperforming GPT-4 and LLaMA in test smell identification. For example, Gemini achieved 89.4\% (Python) and 97.8\% (Java) accuracy for \texttt{Assertion Roulette}, closely aligning with GPT-4's 87.8\% (Python) and 99.1\% (Java), while significantly outperforming LLaMA's 54.9\% (Python) and 82.6\% (Java). Likewise, for \texttt{Constructor Initialization}, Gemini detected 100\% of cases in Python and 76.9\% in Java, surpassing LLaMA’s 100\% (Python) and 75.4\% (Java) and approaching GPT-4’s 100\% (Python) and 83.1\% (Java). These outcomes highlight Gemini’s robustness and consistency in detecting a wide range of test smells.

In contrast, GPT-4 excelled in Java for specific smells such as \texttt{Assertion Roulette} (99.1\%), \texttt{Sleepy Test} (100\%), and \texttt{Constructor Initialization} (83.1\%), outperforming both Gemini and LLaMA in these cases. However, GPT-4's performance in Python was less stable; for instance, for \texttt{Duplicate Assertion}, it detected only 28.4\%, compared to Gemini’s 59.3\% and LLaMA’s 13.5\%.

LLaMA, while demonstrating high performance on simpler smells like \texttt{Empty Test} (100\% in both Python and Java), struggled with more complex smells such as \texttt{Conditional Test Logic}, achieving only 23.7\% (Python) and 64.9\% (Java)—substantially lower than Gemini (85.0\% Python, 98.4\% Java) and GPT-4 (67.5\% Python, 88.1\% Java). These findings indicate that although LLaMA shows strength in specific cases, its overall reliability and coverage are inferior to Gemini and GPT-4.

\medskip
\begin{mdframed}[backgroundcolor=white, linewidth=1pt, linecolor=black]
    \textbf{Observation 1}: Gemini demonstrated the most consistent and robust performance across both Python and Java in terms of test smell detection.
\end{mdframed}
\medskip



Language-specific patterns in test smell detection emerged clearly from our analysis. For Java-specific smells such as \texttt{Mystery Guest} and \texttt{Resource Optimism}, all three LLMs -- GPT-4, Gemini, and LLaMA -- demonstrated similarly high accuracy, indicating their general effectiveness for these cases. However, for more challenging smells like \texttt{Eager Test} and \texttt{Lazy Test}, Gemini consistently outperformed others, achieving 82.8\% and 91.2\% accuracy, respectively, while LLaMA followed with 75.6\% and 66.1\%, and GPT-4 lagged behind.

In contrast, for Python-specific smells, performance varied more. Notably, LLaMA performed best in detecting \texttt{Obscure In-Line Setup} with a 94.1\% accuracy rate, surpassing Gemini (88.2\%) and GPT-4 (82.4\%). However, for \texttt{Suboptimal Assert}, Gemini led with 85.3\%, followed by GPT-4 (58.3\%) and LLaMA (43.8\%), indicating differences in how these models handle nuanced Python testing patterns.

Interestingly, for test smells common to both languages, such as \texttt{Conditional Test Logic} and \texttt{Exception Handling}, detection rates were generally higher in Java than in Python, suggesting that LLMs may be more optimized or better aligned with Java's structure, syntax, and idioms. These findings highlight that while LLMs show strong potential for cross-language test smell detection, their performance can vary depending on the language and the specific smell being analyzed.


\begin{table}[ht]
\centering

\resizebox{0.47\textwidth}{!}{ 
\begin{tabular}{|l|c|c|c|}

\hline
 & \textbf{Gemini} & \textbf{GPT-4}&\textbf{LLaMA}\\
 \hline
\multicolumn{1}{|c|}{\textbf{Smell}} & \textbf{Python / Java}  & \textbf{Python / Java}            & \textbf{Python / Java}            \\ \hline
\textbf{Assertion Roulette}          & 89.4 / \textbf{97.8}     & 87.8 / \textbf{99.1}              & 54.9 / \textbf{82.6}              \\ \hline
\textbf{Conditional Test Logic}      & 85.0 / \textbf{98.4}     & 67.5 / \textbf{88.1}              & 23.7 / \textbf{64.9}              \\ \hline
\textbf{Constructor Initialization}  & \textbf{100.0} / 76.9    & \textbf{100.0} / 83.1             & \textbf{100.0} / 75.4             \\ \hline
\textbf{Duplicate Assertion}         & \textbf{59.3} / 39.7     & 28.4 / \textbf{36.2}              & 13.5 / \textbf{48.3}              \\ \hline
\textbf{Eager Test}                  & - / \textbf{82.8}        & - / \textbf{62.2}                 & - / \textbf{75.6}                 \\ \hline
\textbf{Empty Test}                  & \textbf{100.0} / 84.6    & 50.0 / \textbf{69.2}              & \textbf{100.0} / \textbf{100.0}   \\ \hline
\textbf{Exception Handling}          & 45.1 / \textbf{89.7}     & 44.0 / \textbf{55.3}              & 49.5 / \textbf{90.9}              \\ \hline
\textbf{General Fixture}             & 83.8 / \textbf{87.9}     & 77.8 / \textbf{82.6}              & 64.0 / \textbf{87.9}              \\ \hline
\textbf{Ignored Test}                & \textbf{100.0} / 60.7    & \textbf{100.0} / 56.2             & \textbf{100.0} / 53.9             \\ \hline
\textbf{Lack Cohesion}               & \textbf{31.1} / -        & \textbf{14.0} / -                 & \textbf{4.3} / -                  \\ \hline
\textbf{Lazy Test}                   & - / \textbf{91.2}        & - / \textbf{88.1}                 & - / \textbf{66.1}                 \\ \hline
\textbf{Magic Number Test}           & \textbf{95.9} / 86.9     & \textbf{88.9} / 65.3              & \textbf{53.7} / 53.0              \\ \hline
\textbf{Mystery Guest}               & - / \textbf{97.1}        & - / \textbf{95.2}                 & - / \textbf{96.2}                 \\ \hline
\textbf{Obscure In-Line Setup}       & \textbf{88.2} / -        & \textbf{82.4} / -                 & \textbf{94.1} / -                 \\ \hline
\textbf{Redundant Assertion}         & 25.0 / \textbf{59.2}     & 25.0 / \textbf{59.2}              & \textbf{66.7} / 36.8              \\ \hline
\textbf{Redundant Print}             & \textbf{100.0} / 98.3    & \textbf{100.0} / \textbf{100.0}   & 84.6 / \textbf{100.0}             \\ \hline
\textbf{Resource Optimism}           & - / \textbf{89.3}        & - / \textbf{86.9}                 & - / \textbf{89.3}                 \\ \hline
\textbf{Sensitive Equality}          & - / \textbf{81.6}        & - / \textbf{43.0}                 & - / \textbf{62.1}                 \\ \hline
\textbf{Sleepy Test}                 & \textbf{100.0} / 95.7    & \textbf{100.0} / \textbf{100.0}   & 95.6 / \textbf{98.6}              \\ \hline
\textbf{Suboptimal Assert}           & \textbf{85.3} / -        & \textbf{58.3} / -                 & \textbf{43.8} / -                 \\ \hline
\textbf{Test Maverick}               & \textbf{43.1} / -        & \textbf{13.9} / -                 & \textbf{4.5} / -                  \\ \hline
\textbf{Unknown Test}                & 32.8 / \textbf{58.1}     & 28.7 / \textbf{51.6}              & 22.3 / \textbf{24.7}              \\ \hline

\end{tabular}
}

\caption{Detection percentages for each test smell. The '-' represent that this specific smell does not exist for that language}
\label{tab:det_smell_pct}
\end{table}

\subsection{RQ2: How effective are LLMs in refactoring test code to eliminate identified test smells?} 

\renewcommand{\arraystretch}{0.9}
\begin{table*}[]
\centering
\resizebox{0.8\textwidth}{!}{
\begin{tabular}{|ll|llll|llll|}
\hline
\multicolumn{2}{|c|}{\textbf{Metadata}} & \multicolumn{4}{c|}{\textbf{Smells removed Java}} & \multicolumn{4}{c|}{\textbf{Smells removed Python}} \\ \hline
\multicolumn{1}{|c|}{\textbf{Refactoring smell}} & \multicolumn{1}{c|}{\textbf{Smell}} & \multicolumn{1}{c|}{\textbf{Total}} & \multicolumn{1}{c|}{\textbf{Gemini \%}} & \multicolumn{1}{c|}{\textbf{GPT-4 \%}} & \multicolumn{1}{c|}{\textbf{LLaMA \%}} & \multicolumn{1}{c|}{\textbf{Total}} & \multicolumn{1}{c|}{\textbf{Gemini \%}} & \multicolumn{1}{c|}{\textbf{GPT-4 \%}} & \multicolumn{1}{c|}{\textbf{LLaMA \%}} \\ \hline
\multicolumn{1}{|l|}{Assertion Roulette} & Assertion Roulette & \multicolumn{1}{l|}{874} & \multicolumn{1}{l|}{14.61} & \multicolumn{1}{l|}{16.02} & 10.54 & \multicolumn{1}{l|}{813} & \multicolumn{1}{l|}{26.85} & \multicolumn{1}{l|}{21.65} & 23.51 \\ \hline
\multicolumn{1}{|l|}{Assertion Roulette} & Collateral & \multicolumn{1}{l|}{9984} & \multicolumn{1}{l|}{0.61} & \multicolumn{1}{l|}{0.94} & 1.54 & \multicolumn{1}{l|}{4862} & \multicolumn{1}{l|}{3.01} & \multicolumn{1}{l|}{3.95} & 7.88 \\ \hline
\multicolumn{1}{|l|}{Conditional Test Logic} & Collateral & \multicolumn{1}{l|}{} & \multicolumn{1}{l|}{} & \multicolumn{1}{l|}{} &  & \multicolumn{1}{l|}{5594} & \multicolumn{1}{l|}{-0.43} & \multicolumn{1}{l|}{0.76} & 5.33 \\ \hline
\multicolumn{1}{|l|}{Conditional Test Logic} & Conditional Test Logic & \multicolumn{1}{l|}{266} & \multicolumn{1}{l|}{37.74} & \multicolumn{1}{l|}{40.91} & 26.69 & \multicolumn{1}{l|}{447} & \multicolumn{1}{l|}{24.38} & \multicolumn{1}{l|}{31.17} & 24.27 \\ \hline
\multicolumn{1}{|l|}{Constructor Initialization} & Constructor Initialization & \multicolumn{1}{l|}{29} & \multicolumn{1}{l|}{79.31} & \multicolumn{1}{l|}{10.34} & 65.52 & \multicolumn{1}{l|}{} & \multicolumn{1}{l|}{} & \multicolumn{1}{l|}{} &  \\ \hline
\multicolumn{1}{|l|}{Duplicate Assertion} & Collateral & \multicolumn{1}{l|}{10108} & \multicolumn{1}{l|}{0.63} & \multicolumn{1}{l|}{1.26} & 1.99 & \multicolumn{1}{l|}{6026} & \multicolumn{1}{l|}{1.76} & \multicolumn{1}{l|}{2.92} & 8.2 \\ \hline
\multicolumn{1}{|l|}{Duplicate Assertion} & Duplicate Assertion & \multicolumn{1}{l|}{401} & \multicolumn{1}{l|}{14.11} & \multicolumn{1}{l|}{17.26} & 25.19 & \multicolumn{1}{l|}{194} & \multicolumn{1}{l|}{54.64} & \multicolumn{1}{l|}{73.2} & 78.87 \\ \hline
\multicolumn{1}{|l|}{Eager Test} & Collateral & \multicolumn{1}{l|}{10012} & \multicolumn{1}{l|}{0.64} & \multicolumn{1}{l|}{1.48} & 1.12 & \multicolumn{1}{l|}{} & \multicolumn{1}{l|}{} & \multicolumn{1}{l|}{} &  \\ \hline
\multicolumn{1}{|l|}{Eager Test} & Eager Test & \multicolumn{1}{l|}{860} & \multicolumn{1}{l|}{3.72} & \multicolumn{1}{l|}{3.72} & 5.23 & \multicolumn{1}{l|}{} & \multicolumn{1}{l|}{} & \multicolumn{1}{l|}{} &  \\ \hline
\multicolumn{1}{|l|}{Empty Test} & Collateral & \multicolumn{1}{l|}{} & \multicolumn{1}{l|}{} & \multicolumn{1}{l|}{} &  & \multicolumn{1}{l|}{1484} & \multicolumn{1}{l|}{0.4} & \multicolumn{1}{l|}{0.27} & 0.4 \\ \hline
\multicolumn{1}{|l|}{Empty Test} & Empty Test & \multicolumn{1}{l|}{8} & \multicolumn{1}{l|}{75} & \multicolumn{1}{l|}{100} & 75 & \multicolumn{1}{l|}{2} & \multicolumn{1}{l|}{100} & \multicolumn{1}{l|}{100} & 100 \\ \hline
\multicolumn{1}{|l|}{Exception Handling} & Exception Handling & \multicolumn{1}{l|}{816} & \multicolumn{1}{l|}{7.22} & \multicolumn{1}{l|}{4.18} & 6 & \multicolumn{1}{l|}{91} & \multicolumn{1}{l|}{43.96} & \multicolumn{1}{l|}{54.95} & 54.95 \\ \hline
\multicolumn{1}{|l|}{General Fixture} & Collateral & \multicolumn{1}{l|}{} & \multicolumn{1}{l|}{} & \multicolumn{1}{l|}{} &  & \multicolumn{1}{l|}{5156} & \multicolumn{1}{l|}{3.14} & \multicolumn{1}{l|}{5.43} & 6.98 \\ \hline
\multicolumn{1}{|l|}{General Fixture} & General Fixture & \multicolumn{1}{l|}{70} & \multicolumn{1}{l|}{41.43} & \multicolumn{1}{l|}{55.71} & 60 & \multicolumn{1}{l|}{374} & \multicolumn{1}{l|}{37.7} & \multicolumn{1}{l|}{43.58} & 43.05 \\ \hline
\multicolumn{1}{|l|}{Ignored Test} & Ignored Test & \multicolumn{1}{l|}{40} & \multicolumn{1}{l|}{27.5} & \multicolumn{1}{l|}{25} & 32.5 & \multicolumn{1}{l|}{10} & \multicolumn{1}{l|}{70} & \multicolumn{1}{l|}{80} & 90 \\ \hline
\multicolumn{1}{|l|}{Lack Cohesion} & Collateral & \multicolumn{1}{l|}{} & \multicolumn{1}{l|}{} & \multicolumn{1}{l|}{} &  & \multicolumn{1}{l|}{5966} & \multicolumn{1}{l|}{-6.77} & \multicolumn{1}{l|}{-7.98} & -4.63 \\ \hline
\multicolumn{1}{|l|}{Lack Cohesion} & Lack Cohesion & \multicolumn{1}{l|}{} & \multicolumn{1}{l|}{} & \multicolumn{1}{l|}{} &  & \multicolumn{1}{l|}{235} & \multicolumn{1}{l|}{52.77} & \multicolumn{1}{l|}{56.17} & 63.83 \\ \hline
\multicolumn{1}{|l|}{Lazy Test} & Lazy Test & \multicolumn{1}{l|}{673} & \multicolumn{1}{l|}{5.5} & \multicolumn{1}{l|}{9.83} & 4.76 & \multicolumn{1}{l|}{} & \multicolumn{1}{l|}{} & \multicolumn{1}{l|}{} &  \\ \hline
\multicolumn{1}{|l|}{Magic Number Test} & Collateral & \multicolumn{1}{l|}{8962} & \multicolumn{1}{l|}{0.09} & \multicolumn{1}{l|}{0.34} & 0.99 & \multicolumn{1}{l|}{5000} & \multicolumn{1}{l|}{0.92} & \multicolumn{1}{l|}{1.32} & 6.44 \\ \hline
\multicolumn{1}{|l|}{Magic Number Test} & Magic Number Test & \multicolumn{1}{l|}{} & \multicolumn{1}{l|}{} & \multicolumn{1}{l|}{} &  & \multicolumn{1}{l|}{314} & \multicolumn{1}{l|}{56.05} & \multicolumn{1}{l|}{61.46} & 48.73 \\ \hline
\multicolumn{1}{|l|}{Mystery Guest} & Collateral & \multicolumn{1}{l|}{6660} & \multicolumn{1}{l|}{0.36} & \multicolumn{1}{l|}{0.81} & 0.75 & \multicolumn{1}{l|}{} & \multicolumn{1}{l|}{} & \multicolumn{1}{l|}{} &  \\ \hline
\multicolumn{1}{|l|}{Mystery Guest} & Mystery Guest & \multicolumn{1}{l|}{36} & \multicolumn{1}{l|}{38.89} & \multicolumn{1}{l|}{52.78} & 44.44 & \multicolumn{1}{l|}{} & \multicolumn{1}{l|}{} & \multicolumn{1}{l|}{} &  \\ \hline
\multicolumn{1}{|l|}{Obscure In-Line Setup} & Collateral & \multicolumn{1}{l|}{} & \multicolumn{1}{l|}{} & \multicolumn{1}{l|}{} &  & \multicolumn{1}{l|}{4402} & \multicolumn{1}{l|}{-0.55} & \multicolumn{1}{l|}{-0.55} & -1 \\ \hline
\multicolumn{1}{|l|}{Obscure In-Line Setup} & Obscure In-Line Setup & \multicolumn{1}{l|}{} & \multicolumn{1}{l|}{} & \multicolumn{1}{l|}{} &  & \multicolumn{1}{l|}{17} & \multicolumn{1}{l|}{82.35} & \multicolumn{1}{l|}{94.12} & 82.35 \\ \hline
\multicolumn{1}{|l|}{Redundant Assertion} & Redundant Assertion & \multicolumn{1}{l|}{56} & \multicolumn{1}{l|}{51.79} & \multicolumn{1}{l|}{66.07} & 48.21 & \multicolumn{1}{l|}{4} & \multicolumn{1}{l|}{50} & \multicolumn{1}{l|}{75} & 75 \\ \hline
\multicolumn{1}{|l|}{Redundant Print} & Collateral & \multicolumn{1}{l|}{6244} & \multicolumn{1}{l|}{0.03} & \multicolumn{1}{l|}{0.06} & 0.22 & \multicolumn{1}{l|}{} & \multicolumn{1}{l|}{} & \multicolumn{1}{l|}{} &  \\ \hline
\multicolumn{1}{|l|}{Redundant Print} & Redundant Print & \multicolumn{1}{l|}{45} & \multicolumn{1}{l|}{93.33} & \multicolumn{1}{l|}{93.33} & 86.67 & \multicolumn{1}{l|}{13} & \multicolumn{1}{l|}{84.62} & \multicolumn{1}{l|}{84.62} & 84.62 \\ \hline
\multicolumn{1}{|l|}{Resource Optimism} & Resource Optimism & \multicolumn{1}{l|}{47} & \multicolumn{1}{l|}{14.89} & \multicolumn{1}{l|}{19.15} & 25.53 & \multicolumn{1}{l|}{} & \multicolumn{1}{l|}{} & \multicolumn{1}{l|}{} &  \\ \hline
\multicolumn{1}{|l|}{Sleepy Test} & Collateral & \multicolumn{1}{l|}{6882} & \multicolumn{1}{l|}{0.32} & \multicolumn{1}{l|}{0.58} & 0.64 & \multicolumn{1}{l|}{} & \multicolumn{1}{l|}{} & \multicolumn{1}{l|}{} &  \\ \hline
\multicolumn{1}{|l|}{Sleepy Test} & Sleepy Test & \multicolumn{1}{l|}{50} & \multicolumn{1}{l|}{74} & \multicolumn{1}{l|}{64} & 58 & \multicolumn{1}{l|}{45} & \multicolumn{1}{l|}{84.44} & \multicolumn{1}{l|}{91.11} & 91.11 \\ \hline
\multicolumn{1}{|l|}{Suboptimal Assert} & Collateral & \multicolumn{1}{l|}{} & \multicolumn{1}{l|}{} & \multicolumn{1}{l|}{} &  & \multicolumn{1}{l|}{5192} & \multicolumn{1}{l|}{-0.39} & \multicolumn{1}{l|}{2.08} & 6.86 \\ \hline
\multicolumn{1}{|l|}{Suboptimal Assert} & Suboptimal Assert & \multicolumn{1}{l|}{} & \multicolumn{1}{l|}{} & \multicolumn{1}{l|}{} &  & \multicolumn{1}{l|}{218} & \multicolumn{1}{l|}{84.4} & \multicolumn{1}{l|}{77.52} & 85.32 \\ \hline
\multicolumn{1}{|l|}{Test Maverick} & Collateral & \multicolumn{1}{l|}{} & \multicolumn{1}{l|}{} & \multicolumn{1}{l|}{} &  & \multicolumn{1}{l|}{5448} & \multicolumn{1}{l|}{0.59} & \multicolumn{1}{l|}{1.17} & 4.26 \\ \hline
\multicolumn{1}{|l|}{Test Maverick} & Test Maverick & \multicolumn{1}{l|}{} & \multicolumn{1}{l|}{} & \multicolumn{1}{l|}{} &  & \multicolumn{1}{l|}{202} & \multicolumn{1}{l|}{27.23} & \multicolumn{1}{l|}{30.69} & 32.18 \\ \hline
\multicolumn{1}{|l|}{Unknown Test} & Collateral & \multicolumn{1}{l|}{11010} & \multicolumn{1}{l|}{-0.65} & \multicolumn{1}{l|}{-0.14} & -0.18 & \multicolumn{1}{l|}{5866} & \multicolumn{1}{l|}{-3.17} & \multicolumn{1}{l|}{-4.26} & -2.86 \\ \hline
\multicolumn{1}{|l|}{Unknown Test} & Unknown Test & \multicolumn{1}{l|}{248} & \multicolumn{1}{l|}{28.23} & \multicolumn{1}{l|}{29.72} & 25.4 & \multicolumn{1}{l|}{265} & \multicolumn{1}{l|}{61.13} & \multicolumn{1}{l|}{69.81} & 80 \\ \hline
\end{tabular}
}
\caption{Test smell refactoring percentages for each smell}
\label{tab:smells_removed}
\end{table*}

Table \ref{tab:smells_removed} summarizes the impact of LLM-driven refactoring on various test smells. It is important to note that not all refactoring efforts produced statistically significant effects. The table includes only those cases where changes in test smells were found to be statistically significant, as determined by the Wilcoxon test\cite{Wilcoxon1992}, following a confirmation of data symmetry.

Each column in Table \ref{tab:smells_removed} highlights a critical dimension of our analysis on LLM-based test smell refactoring. The \textit{Refactoring smell} column specifies the target test smell that LLMs were instructed to address during refactoring. The \textit{Smell} column lists the original test smell identified in the test code. Importantly, when the \textit{Smell} column contains "Collateral," it denotes unintended test smells that were either newly introduced or unintentionally removed as side effects of the refactoring process. The columns \textit{Smells removed Java} and \textit{Smells removed Python} present the net changes in the number of test smells for Java and Python projects, respectively. Here, positive values indicate successful removals, reflecting effective refactoring, whereas negative values highlight the accidental introduction of new smells. 

\begin{figure*}[ht]
    \centering
    \includegraphics[width=0.7\textwidth, height=0.6\textheight, keepaspectratio]{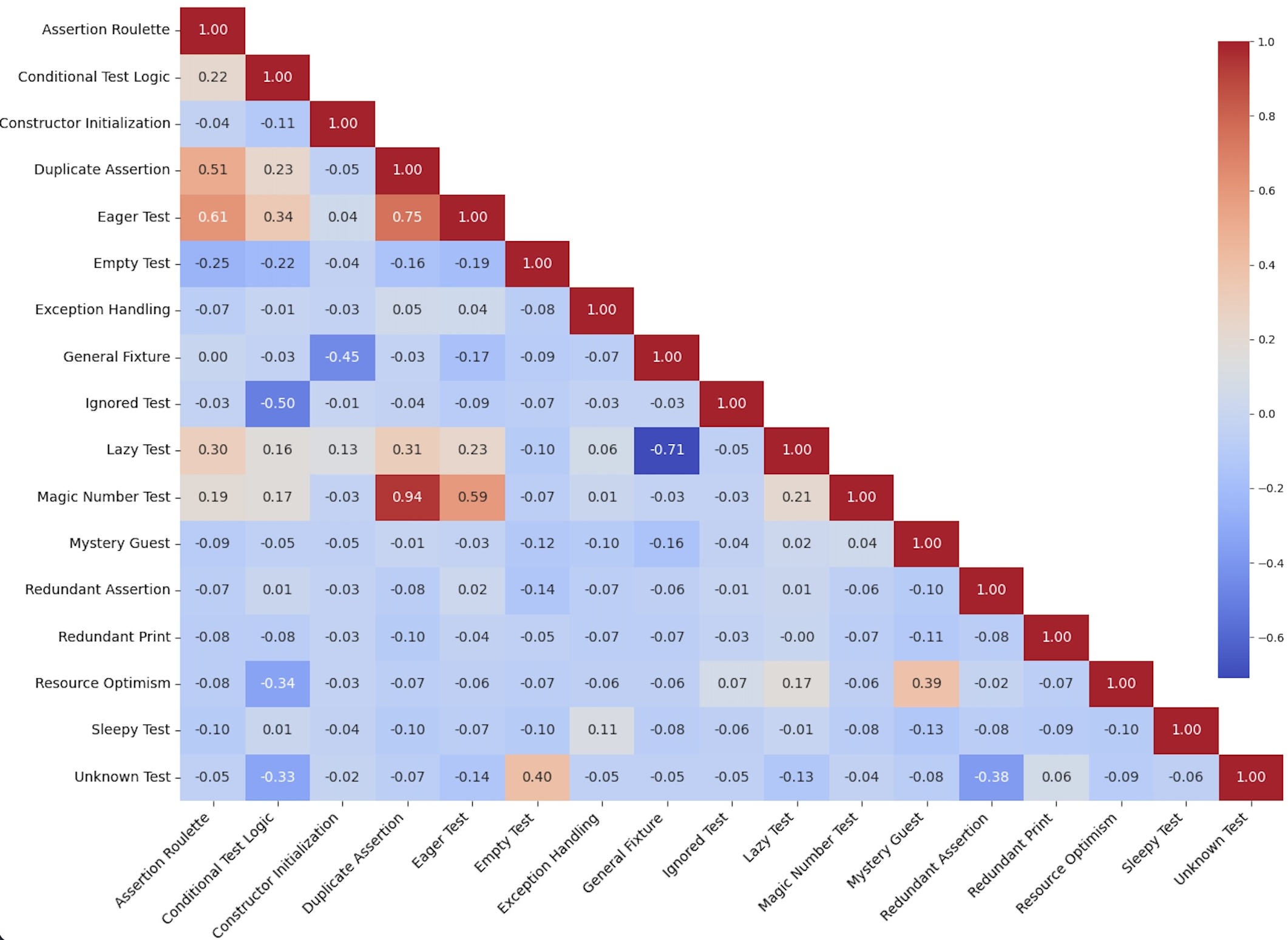}
    \caption{Correlation matrix based on refactoring task.}
    \Description{Correlation matrix based on refactoring task.}
    \label{fig:corr_matrix_java}
\end{figure*}

The study also examined a correlation matrix of all test smells to explore the relationships between test smells and refactoring outcomes. An example of such a matrix is presented in Figure \ref{fig:corr_matrix_java}, showing the removal trends among different test smells.

\subsubsection{\textbf{(RQ2.1)}}

\textbf{Which test smell, if any, is unintentionally removed during refactoring?}

Refactoring aimed at improving code quality and maintainability often leads to the incidental removal of test smells not explicitly targeted, a trend clearly reflected in our findings. The proportional matrices of refactored smells reveal that across all three LLMs, several test smells were almost entirely eliminated, even when not part of the initial refactoring goal. For instance, \texttt{Ignored Test} was consistently removed with a perfect rate of 1.0 across all models and both languages, indicating that LLM-based refactoring inherently addresses such smells as part of general cleanup or standardization. Likewise, \texttt{General Fixture} was completely removed in Java by all LLMs, and \texttt{Constructor Initialization} achieved 100\% removal in Java when using Gemini and GPT-4, regardless of whether it was a refactoring target. Given that both \texttt{General Fixture} and \texttt{Constructor Initialization} concern test setup practices, their simultaneous removal suggests that refactoring efforts impacting setup structures naturally address both smells, highlighting an important side benefit of automated test code improvements.

The unintentional test smell removal rate varied notably across LLMs and programming languages, reflecting differences in their refactoring strategies. For example, \texttt{Duplicate Assertion} was substantially but not fully removed, with proportions like 0.78 for LLaMA (Java), 0.77 for GPT-4 (Java), and 0.89 for Gemini (Java). Similarly, \texttt{Assertion Roulette}, although often a refactoring target, was also partially eliminated without direct instruction, with rates such as 0.53 for LLaMA (Java) and 0.59 for GPT-4 (Python). These results indicate that while LLMs can effectively address related smells during refactoring, the extent of incidental removal depends on both the model and language, underscoring variability in LLM-driven refactoring behaviors.


Conversely, some test smells showed minimal or no unintentional removal, indicating resistance to incidental refactoring. \texttt{Magic Number Test} and \texttt{Sleepy Test} consistently exhibited near-zero removal rates across all models and languages, suggesting that generic refactoring patterns rarely address these smells unless explicitly instructed. Likewise, \texttt{Redundant Assertion} and \texttt{Obscure In-Line Setup} saw little change unless they were direct refactoring targets, highlighting that certain smells require specific interventions beyond general cleanup.


Refactoring one smell can indirectly improve others. For instance, addressing \texttt{Eager Test} often led to partial removal of \texttt{Duplicate Assertion} (0.16) and \texttt{Empty Test} (0.6) in Java when using LLaMA. Likewise, refactoring \texttt{Exception Handling} was frequently associated with reductions in \texttt{Redundant Print} and \texttt{Lazy Test}, especially in Java with GPT-4. These patterns suggest that some test smells share structural relationships that make them responsive to similar refactoring actions.

\medskip
\begin{mdframed}[backgroundcolor=white, linewidth=1pt, linecolor=black]
    \textbf{Observation 2}: Refactoring test smells often led to unintended but beneficial removal of certain test smells.
\end{mdframed}

\subsubsection{\textbf{(RQ2.2)}}

\textbf{Does the removal of test smells unintentionally introduce new ones?}

We also examined test smells that were unintentionally introduced by LLMs during refactoring, revealing notable trade-offs in the refactoring process. Specifically, targeting certain smells, such as \texttt{Conditional Test Logic} and \texttt{Magic Number Test}, often led to the emergence of new, unrelated smells. For example, refactoring \texttt{Conditional Test Logic} frequently introduced \texttt{Lazy Test} and \texttt{Unknown Test}, with introduction rates as high as -0.56 and -0.54 for \texttt{Lazy Test} in Java using LLaMA and Gemini, respectively. Similarly, when refactoring \texttt{Magic Number Test} in Python, \texttt{Suboptimal Assert} was introduced, underscoring how efforts to address one smell can inadvertently create others. A similar pattern was observed when refactoring \texttt{General Fixture} in Java, which led to increased instances of \texttt{Lazy Test} and \texttt{Unknown Test}, with \texttt{Unknown Test} showing introduction rates of -0.5 (GPT-4) and -0.77 (Gemini). Although some smells, like \texttt{Redundant Assertion} and \texttt{Empty Test}, remained largely unaffected, others, such as \texttt{Ignored Test}, occasionally appeared as side effects — for instance, with a -0.1 rate in Python when refactoring \texttt{General Fixture} using Gemini. These findings highlight the complex, interconnected nature of test smells, where addressing one issue may unintentionally give rise to others.

\medskip
\begin{mdframed}[backgroundcolor=white, linewidth=1pt, linecolor=black]
    \textbf{Observation 3}: Refactoring specific test smells often led to the unintended introduction of new smells, highlighting the interdependent and trade-off nature of test smell removal.
\end{mdframed}

\subsubsection{\textbf{(RQ2.3)}}
\textbf{Does the removal of test smells unintentionally impact test coverage?}



The impact of refactoring on code coverage was generally limited, with most refactorings causing minimal change. However, specific cases revealed notable effects. In particular, refactoring the \texttt{Assertion Roulette} test smell led to an average code coverage reduction of 1.76\% across projects, with the most affected cases losing up to 4.71\% coverage. These results indicate that while refactoring can improve code quality by removing test smells, it may also unintentionally weaken the test suite’s ability to comprehensively cover the codebase.



\begin{figure}[ht]
    \centerline{\includegraphics[width=0.5\textwidth]{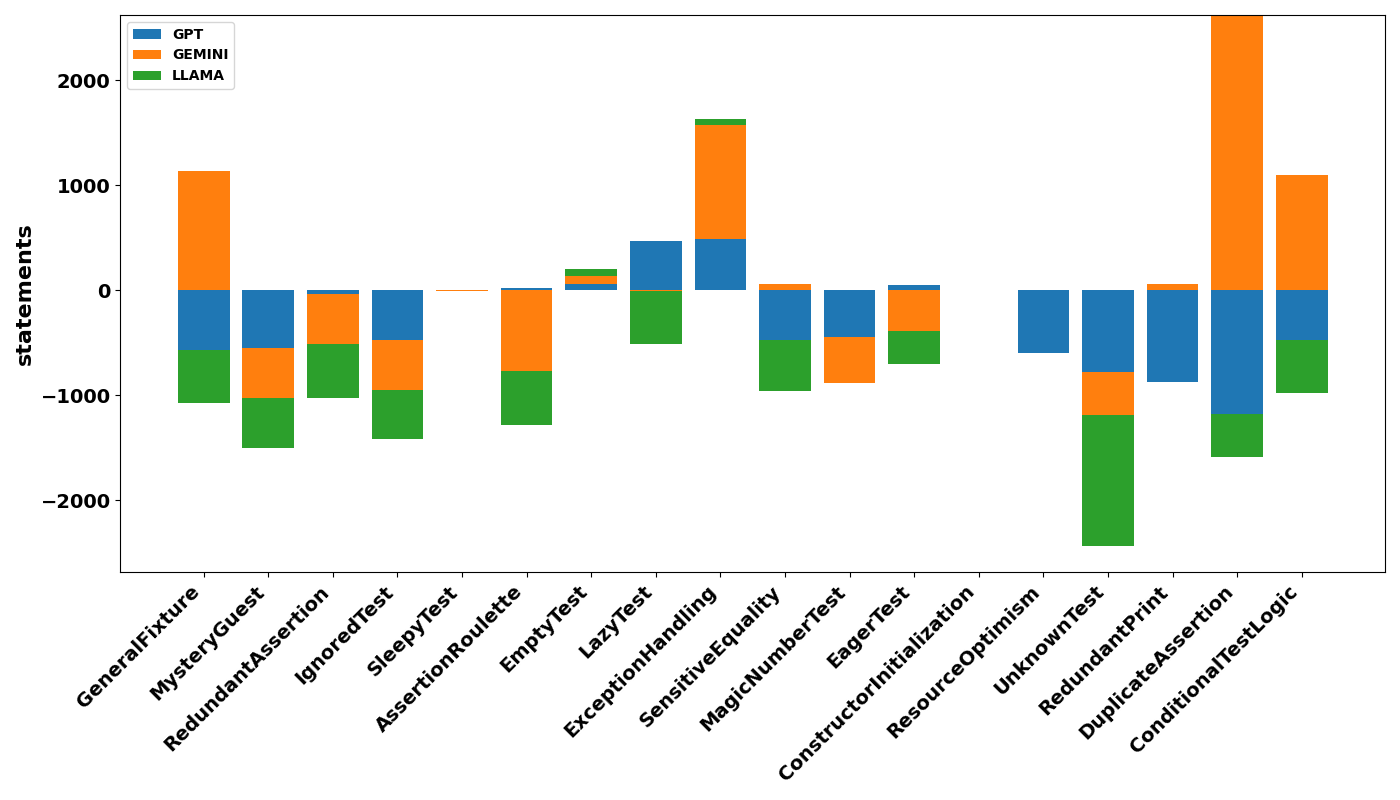}}
    \caption{Statement Coverage Comparison in Java Projects}
    \Description{Statement Coverage Comparison in Java Projects.}
    \label{fig:javaStatements}
\end{figure}

Figure \ref{fig:javaStatements} highlights key differences in how LLMs impact code coverage during refactoring. Gemini consistently produced large positive gains in the number of statements covered for Java projects, notably in smells like \texttt{General Fixture} (1133), \texttt{Exception Handling} (1092), \texttt{Duplicate Assertion} (2624), and \texttt{Conditional Test Logic} (1099), indicating better optimization for coverage. In contrast, GPT-4 and LLaMA often showed negative coverage impacts, reducing the number of covered statements, for example, in \texttt{General Fixture} (-574, -501), \texttt{Sensitive Equality} (-476, -483), and \texttt{Duplicate Assertion} (-1175, -416). An exception was observed for \texttt{Lazy Test}, where GPT-4 improved coverage (468), but Gemini and LLaMA did not. These patterns suggest that while Gemini tends to improve coverage, GPT-4 and LLaMA may introduce refactorings that inadvertently reduce it. Additionally, some smells like \texttt{Constructor Initialization} showed no effect on coverage across all models, indicating either no refactoring occurred or changes had no coverage impact.

\begin{figure}[ht]
    \centerline{\includegraphics[width=0.5\textwidth]{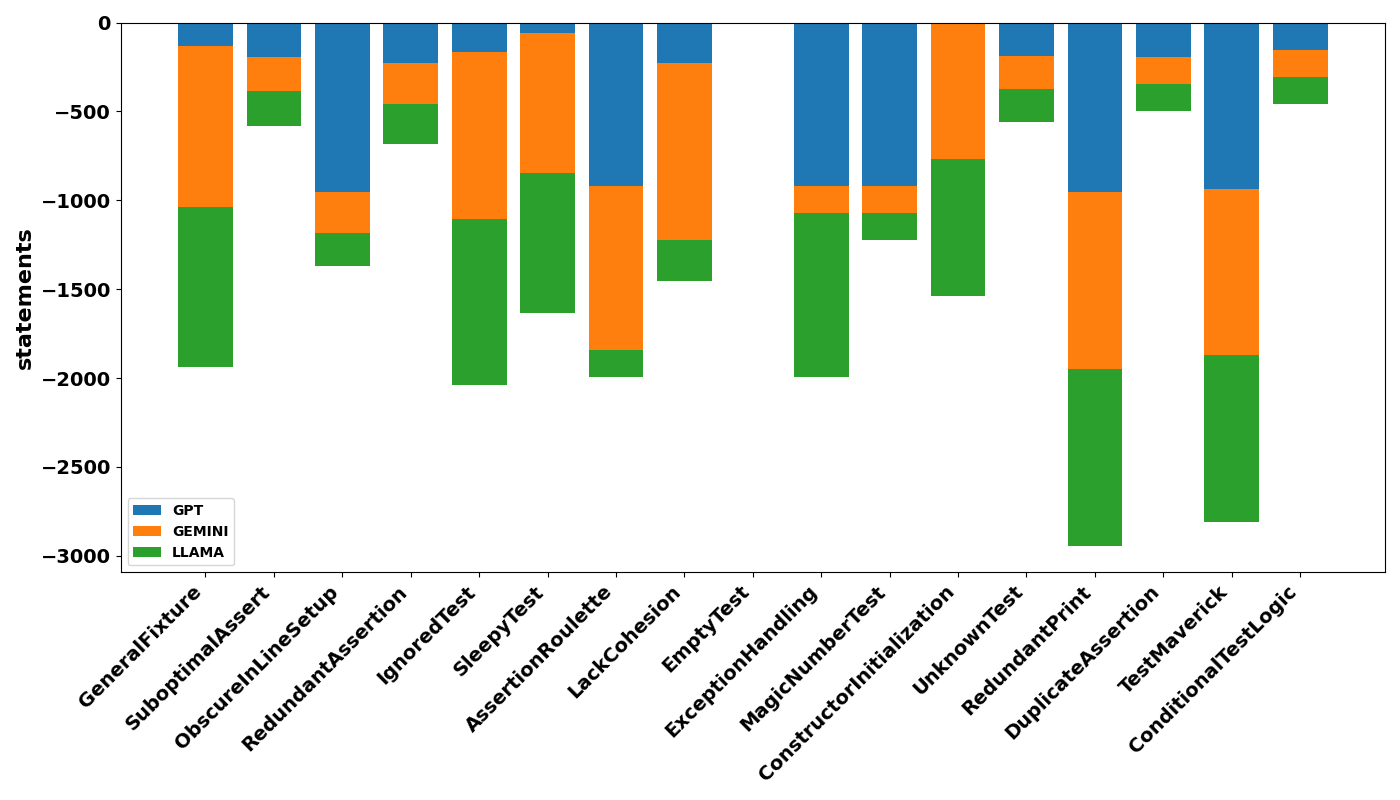}}
    \caption{Statement Coverage Comparison in Python Projects}
    \Description{Statement Coverage Comparison in Python Projects}
    \label{fig:pythonStatements}
\end{figure}

A predominantly negative trend is evident in Python projects (Figure \ref{fig:pythonStatements}), where all three models—Gemini, GPT-4, and LLaMA--generally reduced code coverage after refactoring. Several test smells, including \texttt{Suboptimal Assert} (-194), \texttt{Redundant Assertion} (-228), \texttt{Unknown Test} (-186), \texttt{Test Maverick} (-936), and \texttt{Conditional TestLogic} (-152), showed identical negative impacts across models, indicating shared refactoring behaviors. Notably, \texttt{Redundant Print} (-954) and \texttt{Test Maverick} (-936) experienced some of the largest coverage drops. In contrast, smells like \texttt{Empty Test} showed no coverage change, suggesting limited or absent refactoring in these cases.


We analyzed the change in uncovered statements between refactored and original code, where positive values indicate more uncovered statements (a negative impact), and negative values indicate fewer uncovered statements (a positive impact).

\begin{figure}[ht]
    \centerline{\includegraphics[width=0.5\textwidth]{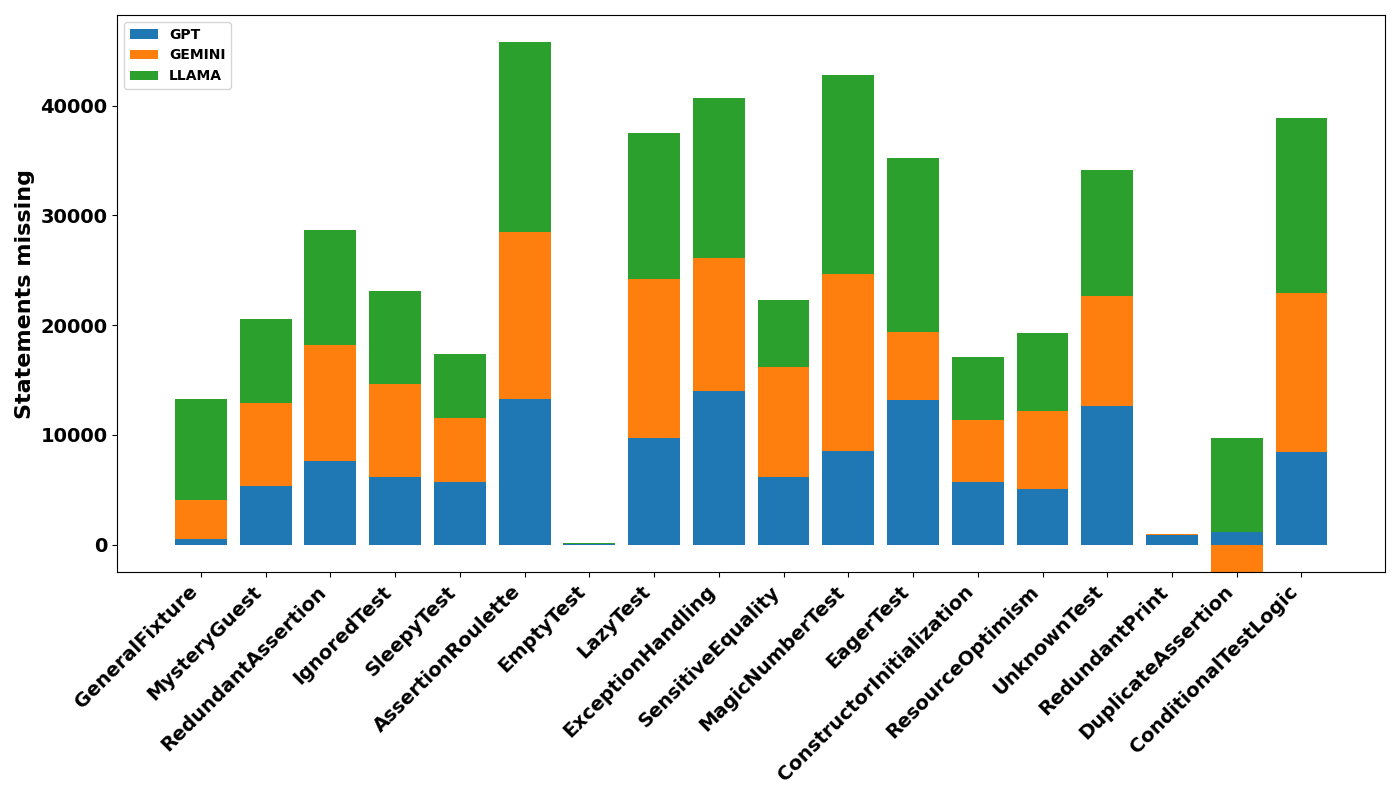}}
    \caption{Statements missing coverage in Java projects.}
    \Description{Statements missing coverage in Java.}
    \label{fig:javaMissing}
\end{figure}

For Java projects (Figure \ref{fig:javaMissing}), all three LLMs frequently increased the number of uncovered statements, signaling coverage degradation in many cases. Notably, \texttt{Magic Number Test} and \texttt{Assertion Roulette} were among the most impacted. Gemini often showed higher increases than GPT-4, for instance, in \texttt{Magic Number Test} (16,068 vs. 8,578) and \texttt{Lazy Test} (14,463 vs. 9,732)—indicating a stronger negative effect. LLaMA generally performed worse than both, with even higher uncovered statement counts, such as in \texttt{Magic Number Test} (18,126) and \texttt{Redundant Print} (40 vs. 0). These results suggest that LLaMA had the most negative impact on code coverage, followed by Gemini and GPT-4. Furthermore, substantial variation across models highlights inconsistent refactoring behaviors and their uneven effects on coverage.

\begin{figure}[ht]
    \centerline{\includegraphics[width=0.5\textwidth]{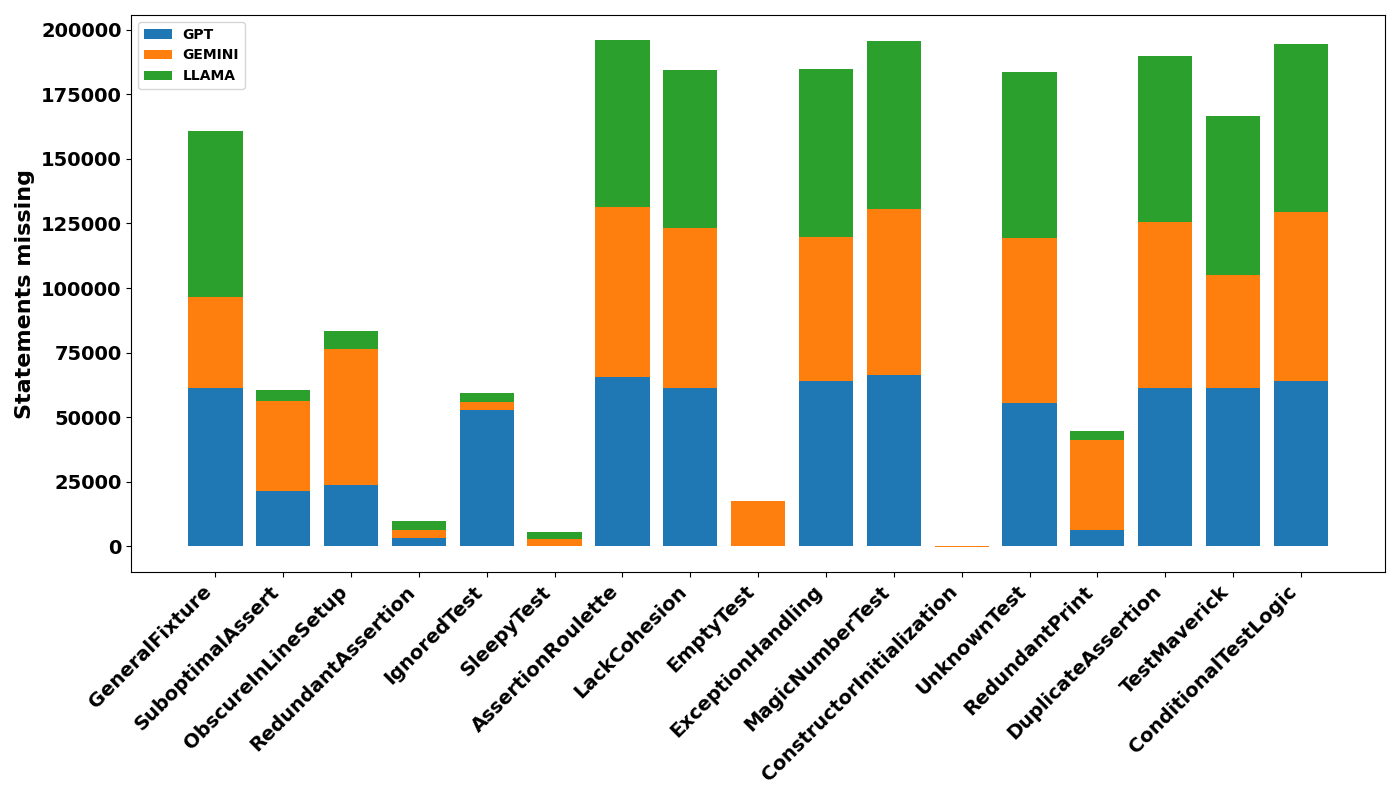}}
    \caption{Statements missing coverage in Python projects.}
    \Description{Statements missing coverage in Python projects.}
    \label{fig:pythonMissing}
\end{figure}

For Python projects (Figure \ref{fig:pythonMissing}), all models showed a significant increase in uncovered statements, with Gemini performing the worst. Gemini led to substantial coverage losses in cases like \texttt{Assertion Roulette} (65,868) and \texttt{Conditional Test Logic} (65,118), along with severe impacts on \texttt{Suboptimal Assert} (34,898), \texttt{Obscure In-line Setup} (52,568), and others. LLaMA also performed poorly, with high uncovered statements such as 64,360 for \texttt{Assertion Roulette} and around 64,000 for both \texttt{Magic Number Test} and \texttt{Exception Handling}. Although GPT-4 and Gemini showed slightly more consistent outcomes compared to LLaMA, overall, the results highlight substantial and inconsistent coverage degradation across all models during Python test smell refactoring.



\medskip
\begin{mdframed}[backgroundcolor=white, linewidth=1pt, linecolor=black]
    \textbf{Observation 4}: Refactoring test smells led to inconsistent and often negative impacts on code coverage across both Java and Python projects, with Gemini and LLaMA frequently introducing more uncovered statements than GPT-4.
\end{mdframed}
\medskip
\section{Discussion}

\label{section:discussion}

In this section, we discuss the implications of our study’s results for practice and research.

\subsection{Implications for Practice}
Our findings demonstrate that LLMs show real potential for detecting and refactoring some test smells and present significant drawbacks for those use cases. Both the good and the bad parts of our findings can impact software engineering in substantial ways:

\textbf{Incorporate LLM powered test smell detection and refactoring in IDEs.} Although for some earlier adopters in the software development industry, integrating LLMs into the software engineer normal workflow is already a reality \cite{agarwal2024copilot, tang2024towards}, incorporating LLM powered test smell detector and refactoring assistant on IDEs might present issues if the models selected were not carefully selected and quality assured. This is a significant concern because our findings indicate that each language model presents vastly different performances on detecting and refactoring for each kind of test smell, and because of this detail changing the underlying model being used can suddenly completely change the quality of the tool. A factor that might aggravate this issue is that the LLM space is known for constantly launching new and differently capable language models constantly, each one with its specific characteristics. This extra factor increases the risk of this problem because keeping track of what kind of model and its specific version is a difficult task.
Although the significant disparity in LLMs performances can present issues, developing a quality assurance dataset and tools for periodically test and record the performances in each software engineering task should be enough to catch and deal with this problem before it impacts real world developers.

\textbf{Integrating LLMs into CI/CD pipelines.} Another direction is to incorporate these automated detection and refactoring steps into CI/CD pipelines \cite{chen2024challenges}. However, our data show that while LLMs successfully refactor some smells, the output of these models can introduce significant unpredictability in applications that rely on it for correct and repeatable output, even sometimes introducing new test smells, or reduce code coverage. This problem was observed for all LLMs we tested, in all prompt templates used in the study. Given this context, our results suggest that software developers should preferably refrain from integrating those language models where precise, predictable and repeatable output is a requirement. 
If this kind of use case for LLM is unavoidable, this problem can be addressed by using the solution described in the practice implication 3: deterministic, purpose specific detection and refactoring tools. These tools provide a reproducible but not context specific output that can be used to compensate for the drawbacks of LLMs while still allowing them to enjoy their advantages.

\textbf{Using LLMs to detect and refactor smelly test code.} Building on the potential of LLMs for test smell detection and automated refactoring, it is crucial to address their unpredictability and variability in output. While LLMs are often considered as potential substitutes for existing solutions, their inconsistent behavior presents challenges, particularly in tasks that require reliability.
A possible way to mitigate these drawbacks is by integrating LLMs with a deterministic, context-independent tool \cite{10795061}. This complementary tool would provide stable, consistent outputs in scenarios where contextual understanding is not required, while the LLM could focus on delivering richer, more adaptive suggestions when deeper insights are necessary.
Such a hybrid approach would allow teams to leverage the strengths of both methods. For instance, deterministic techniques can reliably identify common test smells—such as duplicated assertions or unnecessarily long test cases—without the need for contextual interpretation. Meanwhile, LLMs can enhance the process by generating in-depth explanations and tailored refactoring recommendations.
By distributing tasks in this manner, organizations can maximize the benefits of LLM-driven insights while mitigating their limitations, ultimately leading to more reliable and higher-quality software testing practices.

\subsection{Implications for Research}
Our results suggest multiple future research directions on taking advantage of LLMs potential while addressing their current limitations.

\textbf{Fine-tuning models for test-related software engineering tasks}. Fine-tuning existing models on domain-specific data—such as the testing frameworks, conventions, or build tools unique to an organization—has the potential to substantially improve both detection accuracy and refactoring reliability \cite{An_Empirical_Comparison}. By training an LLM to recognize and adapt to specific coding patterns, environment constraints, and testing norms, researchers can reduce false positives and ensure that the suggested refactorings align with established best practices. Additionally, targeted fine-tuning may allow models to handle specialized edge cases, such as company-specific libraries or workflows, resulting in more trustworthy and context-aware test smell detection. Future studies can explore various fine-tuning methods (e.g., parameter-efficient or multi-task learning approaches) to balance model performance with the computational and maintenance costs associated with constantly evolving project environments.

\textbf{LLM Agents for software engineering}. Agent-based systems offer a more dynamic approach by enabling LLMs to iteratively collaborate with external tools \cite{he2024llm}, such as coverage analyzers, static checkers, and test runners. In this setup, the LLM can propose refactorings and then solicit immediate feedback from these external analyzers—evaluating metrics such as test coverage, code complexity, and style compliance—before deciding on the next step. This interactive cycle continues until all identified issues have been resolved or no new problems arise. Such an agent framework provides greater robustness: the LLM can correct its own oversights in real time, and the final output is thoroughly vetted against automated checks. Researchers could further investigate ways to structure these interactions, including the optimal sequence of feedback loops, confidence thresholds, or fallback strategies, to maximize the impact of LLM-based refactoring without creating additional maintenance overhead \cite{dos2015autorefactoring}.

\textbf{Knowledge distillation and small powerful LLMs.} While large-scale models offer substantial predictive power, deploying them in enterprise settings or resource-constrained environments can be challenging. Model distillation addresses this tension by transferring the knowledge learned by large LLMs into smaller, specialized models \cite{liu2024deepseek, guo2025deepseek}. This enables organizations to integrate distilled models more seamlessly into continuous integration (CI) pipelines or shared development environments—particularly relevant for large teams or companies with stringent memory and computational limits. By focusing on test smells and refactoring tasks, the distilled models can be trained to specialize in essential detection patterns and correction heuristics. Researchers might explore various distillation strategies—such as teacher–student frameworks, multi-phase training, or architecture search—to strike an optimal balance between performance, size, and adaptability.

\textbf{Real-world validation for LLM-based test smell detection and refactoring}. Ultimately, examining how LLM-driven test smell detection and refactoring interact with real-world software development is crucial. Practical validation strategies include in-depth case studies \cite{runeson2012case} where researchers embed these tools within actual project teams, observing day-to-day usage, adoption hurdles, and the measurable impact on code quality. Such studies can uncover hidden barriers (e.g., lack of trust in AI-generated suggestions or conflicts with existing team processes) or unanticipated advantages (e.g., reduced onboarding time for junior engineers). Broader surveys and semi-structured interviews \cite{seaman1999qualitative} with practitioners also provide an avenue to gather diverse perspectives, shedding light on subjective elements like interpretability, trust, and mental effort. These findings not only help refine the tools to be more user-friendly and context-aware, but also inform guidelines and best practices for successful LLM integration in the software testing lifecycle.

\section{Threats to Validity}

\label{section:threats_to_validity}

\textbf{Construct Validity.} The accuracy of test smell definitions and the precision of detection and refactoring methods present a potential threat to construct validity. Our study relies on PyNose for smell detection and on carefully crafted prompts to guide GPT-4, making outcomes dependent on both the tools' effectiveness and the clarity of task formulation. To mitigate this threat, we grounded our test smell definitions in well-established literature and adopted an iterative prompt validation process to ensure that both smell detection and refactoring tasks were accurately represented and understood by the model~\cite{ahmed_2021}.


\textbf{External Validity.} The broader relevance of our findings may be limited by focusing on open-source projects, as open-source projects often follow specific coding standards and testing practices that may not generalize to the broader ecosystem or other languages. We selected projects previously validated in past studies to mitigate this, ensuring a solid foundation for our analysis. Additionally, we carefully documented our selection criteria and project characteristics, allowing future replication of our work~\cite{LLMsTestSmellsSupp}.


\textbf{Conclusion Validity.} The validity of our conclusions depends on the assumption that improvements or stability in observed metrics after refactoring reflect real enhancements in code quality. However, we acknowledge that such metrics may not capture all aspects of quality and maintainability. To mitigate this, we complemented quantitative metrics with qualitative analysis when possible, offering a more well-rounded evaluation of refactoring outcomes.


\section{Conclusion and Future Work}

\label{section:conclusion_and_future_work}

This study highlights key considerations for using LLMs in software testing. While models like Gemini effectively detect test smells through prompt-based instructions, their standalone refactoring often compromises test coverage. The consistent coverage drop across all models suggests the need for alternative approaches, especially for Java test suites. A promising direction is leveraging multi-agent conversational systems, where multiple specialized LLM agents—such as a ``test smell detector,'' a ``refactoring agent,'' and a ``test coverage checker'' --collaborate to review, propose, and validate changes. Such systems can simulate developer collaboration, allowing agents to critique and refine each other's suggestions, leading to more accurate and context-aware refactorings that are less likely to harm coverage or functionality. This layered interaction may address the limitations of single-shot, question-answering models by creating iterative feedback loops that better align with real-world developer workflows. Additionally, our findings show that providing LLMs with more detailed information about each test’s purpose and its connection to production code improves refactoring outcomes. This suggests an important research opportunity to explore which specific types of context (e.g., method documentation, code comments, intended test behaviors) most enhance LLM performance, and how these can be effectively incorporated for higher quality refactoring.

\bibliographystyle{IEEEtran}

\bibliography{references}

\end{document}